\documentclass[iop,numberedappendix]{emulateapj}
\usepackage[bookmarks,colorlinks,citecolor=cyan]{hyperref}
\usepackage{ulem}
\usepackage{breakurl}
\usepackage{refcheck}

\def\ngc#1{\hbox{NGC\,#1}}
\def\feh{\hbox{\rm [Fe/H]}}

\def\hst{{\it HST\/}}
\def\teff{\hbox{$\rm T_{eff}$}}

\bibpunct[; ]{(}{)}{;}{a}{}{,}
\defcitealias{Fer98}{F98}
\nocitenames
\norefnames

\slugcomment{}

\shorttitle{HB stars temperature distribution}
\shortauthors{E.\,P. Lagioia et al.}

\begin{document}
\title{The temperature distribution of Horizontal Branch stars:
methods and first results.\altaffilmark{1}}
\author{
 E.~P.~Lagioia\altaffilmark{2},
 E.~Dalessandro\altaffilmark{2},
 F.~R.~Ferraro\altaffilmark{2},
 M.~Salaris\altaffilmark{3},	
 B.~Lanzoni\altaffilmark{2},
 A.~Pietrinferni\altaffilmark{4},
 S.~Cassisi\altaffilmark{4}
 }

\altaffiltext{1}{Based on on observations with the NASA / ESA \hst, obtained at the
Space Telescope Science Institute, which is operated by AURA, Inc., under NASA
contract NAS5-26555.}
\altaffiltext{2}{Dipartimento di Fisica e Astronomia, Universit\`a degli Studi
di Bologna, via Ranzani 1, I-40126, Bologna, Italy; email:
\email{edoardo.lagioia2@unibo.it}} 
\altaffiltext{3}{Astrophysics Research Institute, Liverpool John Moores
University, IC2 Liverpool Science Park, 146 Brownlow Hill, L3 5RF, Liverpool, UK}
\altaffiltext{4}{INAF - Osservatorio Astronomico di Collurania, via Mentore 
Maggini, I-64100, Teramo, Italy}


\begin{abstract}
As part of a large project aimed at characterizing the ultraviolet (UV)
properties of globular clusters, we present here a theoretical and observational
analysis aimed at setting the framework for the determination of horizontal
branch (HB) temperature distributions. Indeed this is a crucial
information to understand the physical parameters shaping the HB morphology in
globular clusters and to interpret the UV emission from unresolved stellar
systems. We found that the use of zero age HB color - \teff\ relations is a
robust way to derive effective temperatures of individual HB stars.  We
investigated the most suitable colors for temperature estimates, and the effect
on the color - \teff\ relations of variations of the initial chemical
composition, and of the evolution off the zero age horizontal branch.  As a test
case, we applied our color - \teff\ calibrations to the Galactic globular
cluster M15. The photometry of M15 has been obtained with the Wide Field and
Planetary Camera 2 on board the \hst. The HB of M15 turned out to have a
multimodal distribution, with a main component peaked at $\teff \sim 8,000$\,K
and confined below $\teff \sim 10,000$\,K. The second component is peaked at
$\teff \sim 14,000$\,K  and extends up to $\teff \sim 20,000$\,K.  The vast
majority ($\sim 95\%$) of the HB stars in M15 is below 20,000\,K, in agreement
with the lack of a well populated extreme HB observed in other metal-poor
globular clusters. We also verified that the temperatures derived with our
analysis are consistent with spectroscopic estimates available in the
literature.
\end{abstract}


\section{Introduction}
The ultraviolet (UV) flux of old stellar systems like Galactic globular clusters
(GGCs) is dominated by a minority of objects. Amongst them, blue Horizontal
Branch (BHB) stars and their progeny, like post-early Asymptotic Giant Branch
(post-EAGB) or AGB-manqu\'{e} stars, are the strongest UV emitters
\citep{Wel72}. The relative contribution of the various types of stars to the
total UV emission, as well as the factors that may lead to larger or smaller
populations of UV-bright stars in a cluster still remain open issues
\citep{Gre90,Cas91,Dor95,Lee02,Ric05,Soh06}. A complete census of hot stars in
stellar populations is therefore a prerequisite for comparing evolutionary
models with observations \citep{Dal12,Sch12}. In this regard GGCs are of
paramount importance, since they are the closest example in nature of relatively
simple systems and span a large range in metallicity, a small range in age and
perhaps a range in helium abundance. Moreover, being typically populated by some
$10^5$ stars, even short-lived evolutionary phases can be properly sampled. GGCs
can therefore be used to test stellar evolution models, one of the basic
ingredients for the interpretation of the integrated light from distant
galaxies.

Our ability to predict the fractions of UV emitters in clusters is deeply linked
to understanding the physical mechanisms driving the Horizontal Branch (HB)
morphology. It is commonly accepted that metallicity is the first parameter
affecting the HB morphology: metal-rich GGCs typically have red HBs, while
metal-poor ones have more extended and bluer HBs. However, there are several
clusters with the same metallicity showing remarkable differences in HB
morphology. Therefore, metallicity alone is not able to explain the complex HB
zoology in GGCs \citep{Fre81}. This issue, known as the ``2nd parameter
problem'', has attracted the attention of several authors in the last decades
\citep{San67,vdB67,Fus93,Lee94,Cat09,Dot10,Gra10,Mil14}. Although there is a
general consensus about the fact that age is the main {\it global} 2nd
parameter, i.e. a parameter that varies from cluster to cluster
\citep{Dot10,Gra10}, no combination of metallicity and age can fully account for
the different HB morphology observed in several GGCs. In this sense a clear
example is given by the clusters M3, M13, M79 and M80
\citep{Fer97-2,Fer98,Dal13-1}. The necessity of an additional parameter was
recently advised by \cite{Dot10} and \citet[][see also
\citealt{Dan05,Dal11,Dal13-1,Mil14}]{Gra10} who suggested the cluster density
\citep[see also][]{Fus93} and an internal spread of He abundance, respectively,
as possible HB third parameters\,\footnote{It is important to recall that also
mass loss efficiency along the Red Giant Branch (RGB) plays a crucial role in
determining the mass of the star along the Zero Age Horizontal Branch (ZAHB),
with a clear impact on the HB morphology \citep[see][]{Cat00,Ori07,Ori14}.},
that vary among different sub-populations within the same cluster. Indeed
\citet[][see also \citealt{Dan05}]{Dal11} have been able to reproduce the
complex HB morphology of the massive GGC \ngc2808 by assuming different He
abundances for the three sub-populations revealed by photometric \citep{Pio07}
and spectroscopic \citep{Bra10,Pas11} analyzes. In addition, \cite{Dal13-1} have
shown that variations in the He abundance can account for the differences
observed among M3, M13 and M79.

The typical approach to the study the HB morphology is to make use of parameters
related to physical properties of HB stars
\citep{Fer93,Fus93,Lee94,Buo97,Dot10,Mil14}. The effective temperature
distribution of HB stars is the most efficient way to describe the HB
morphology. It can be used to constrain the parameters driving the HB
morphology in globular clusters thus shedding new light on our present
understanding of the ``2nd parameter problem''. In addition the temperature
distribution of HB stars is a prime ingredient to interpret the UV emission from
unresolved stellar systems.  The HB star effective temperature distribution has
been studied in GGCs by means of optical data \citep[e.g.,
][]{Rec06,Mon12,Sal13}. However, in the optical and for $\teff > 12,000$ -
15,000\,K, HB stars get increasingly faint and describe an almost vertical
sequence (at approximately constant colors) because of the strong increase of
the bolometric corrections with \teff. This makes color variations weakly
sensitive to changes in temperature (a change of a few tenths in color
corresponds to a variation of several thousands degrees in \teff). Indeed, for
GGCs with extremely BHBs, effective temperatures derived from optical
color-magnitude diagrams (CMDs) can be underestimated by up to $\sim 10,000$\,K,
as shown in the spectroscopic \citep{Moh04} and photometric \citep{Dal11}
analysis of the HB temperature distribution of NGC2808. 

To overcome this problem, we have surveyed 31 GGCs spanning a wide range of
metallicities, mass and structural parameters with the Wide Field and Planetary
Camera 2 (WFPC2) on board the Hubble Space Telescope (\hst), by using a
combination of UV and optical filters (Prop. 11975, PI: Ferraro).  This dataset
is ideal to characterize the properties of exotic objects in GGCs
\citep{Fer01,Fer09} and indeed some results from this survey aimed mainly at
characterizing the properties of Blue Straggler Stars (BSSs) have been recently
published \citep{Con12,Fer12,San12,Dal14,San14}. With only a few exceptions, all
target clusters have been observed in the F170W, F255W, F336W and F555W bands.
This filter setup allows us to derive temperatures in the most appropriate CMD
over the entire extension of the HB. The dataset is also complemented with
observations collected during the last 20 years by our group with the same
camera
\citep{Fer97-1,Fer97-2,Fer98,Fer03,Lan07-1,Lan07-2,Lan07-3,Dal08,Dal11,Dal13-1,Dal13-2}.

In this work we first introduce our methodology to determine the \teff\
distribution along the HB of our sample of GGCs, and discuss several possible
sources of systematic errors. 

We derive here the \teff\ distribution of HB stars in the massive GGC \ngc7078
(M15). Methods and results presented in this work will be used in the
following papers of the series. We note that we chose this cluster as test case
for our methods because it is the only target of the entire sample with a blue
HB observed in the two far UV filters, namely F160BW and F170W. This will allow
us to make an homogeneous comparison with previous results obtained only in the
F160BW filter \citep{Fer98,Dal11,Dal13-1}.

In $\S2$ we describe the approach used for the \teff\ estimate. $\S3$ describes
the observations and data reduction procedures. In $\S4$ we report the criteria
adopted for the HB stars selection.  In $\S5$ the HB temperature distributions
is derived. $\S5$ summarizes the main results.

\section{Derivation of the effective temperatures}\label{sec:efftemp}
Our methodology to determine the \teff\ distribution of the observed HB stars is
very similar to what employed in the study of \ngc2808 HB by \cite{Dal11}. We
consider the observed colors of each individual HB star in a given cluster, and
derive the \teff\ by interpolating a cubic spline along the color - \teff\
relation given by the appropriate theoretical ZAHB, suitably reddened according
to the cluster estimated extinction. This approach neglects the post-ZAHB
evolutionary effects on the color - \teff\ relation, that  we will assess later
in this section.

\begin{figure*} 
\centering
\includegraphics[width=\textwidth]{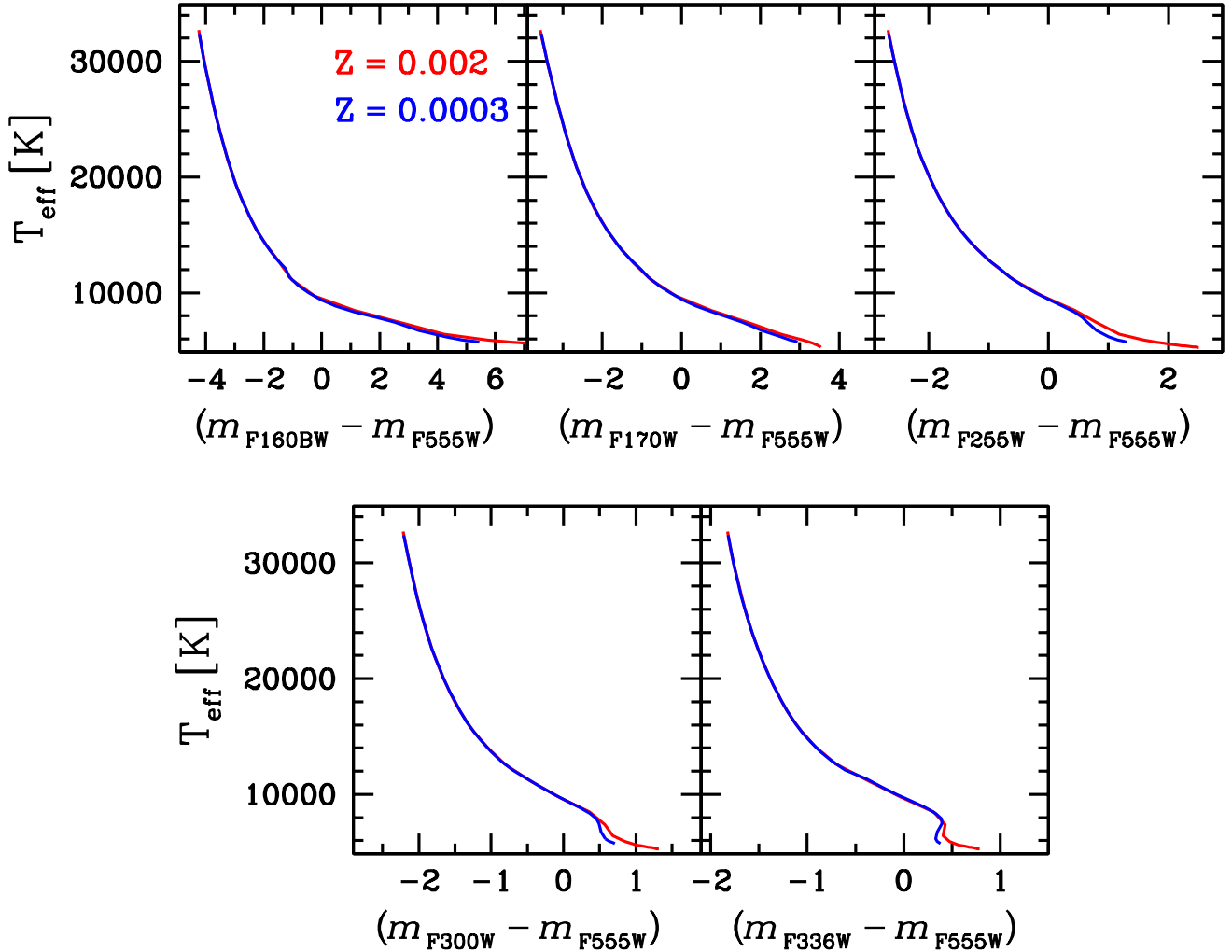} 
\caption{Color - \teff\ relations from ZAHB models for several color
combinations obtained from our photometry, and two different metallicities (see
text for details).} 
\label{fig:col-teffZ} 
\end{figure*}

\begin{figure*} 
\centering
\includegraphics[width=\textwidth]{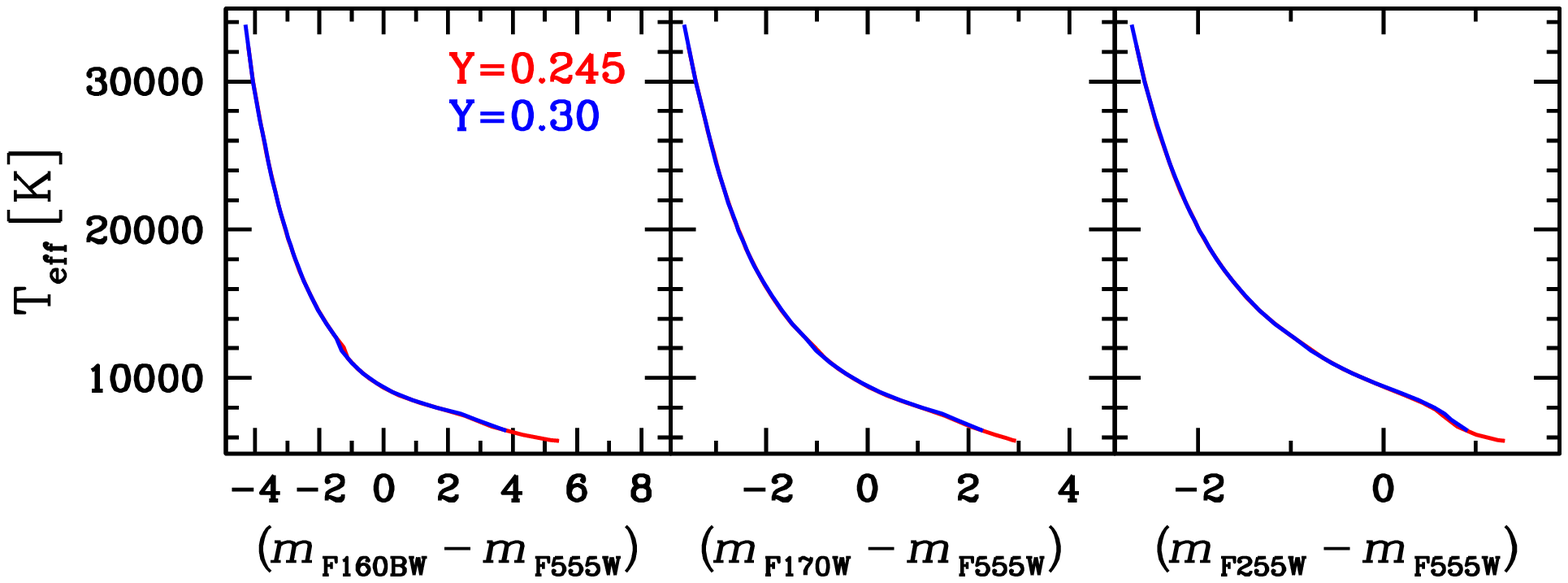} 
\caption{As Fig.~\ref{fig:col-teffZ} but for the effect of the initial He mass
fraction Y, for the three colors we will use in our study (see text for
details).} 
\label{fig:col-teffY}
\end{figure*}

We have employed the
BaSTI\,\footnote{\url{http://basti.oa-teramo.inaf.it/index.html}}
$\alpha$-enhanced models by \cite{Pie06}; to account for the effect of radiative
levitation in the stellar atmospheres, that sets in at about 11,500 - 12,000\,K
\citep[see for example][]{Pac06}, we have applied bolometric corrections
appropriate for [Fe/H]=0.0 and scaled-solar mixture to the HB models, when
\teff\ is above 12,000\,K~\footnote{We tested the use of a super-solar
abundance, i.e. $\feh = +0.5$, for the bolometric corrections at \teff\ larger
than 12,000K and we found no appreciable difference in the resulting color -
\teff\ relation.}. While this is a crude approximation (made necessary by the
lack of more suitable complete grids of HB stellar evolution and atmosphere
models), it still provides reliable measures of the temperatures when compared
to spectroscopic estimates for hot HB stars \citep[see discussion in][]{Dal11}.
Remarkably, we found that the color - \teff\ relation of our chosen filter
combinations (but not the individual magnitudes),  is practically identical to
the counterpart obtained from ZAHB models with the same composition but without
the recipe to mimic the effect of radiative levitation above 12,000\,K.
Figure~\ref{fig:col-teffZ} displays the ZAHB color - \teff\ relations for five
filter combinations (employing the filters available from our observations) with
sufficiently large baseline and different metallicities.  The effect of the
metals is visible only below $\sim 8,000$\,K, mainly for ($m_{\rm F255W} -
m_{\rm F555W}$) and for the redder filter combinations that we are not going to
use in our analysis (see below). Moreover, the differences shown in Fig.
\ref{fig:col-teffZ} are upper limits, since they are obtained for extremely
different metallicities. Our model grid allows us to pick ZAHB sequences with a
metallicity within less than a factor of two of the estimates for each surveyed
cluster, and therefore the related uncertainty on the individual \teff\ values
below 8,000\,K is always kept below $\sim 100$\,K. We can conclude that
metallicity does not play a major in the determination of the color - \teff\
relations.

\begin{figure} 
\centering
\includegraphics[width=\columnwidth]{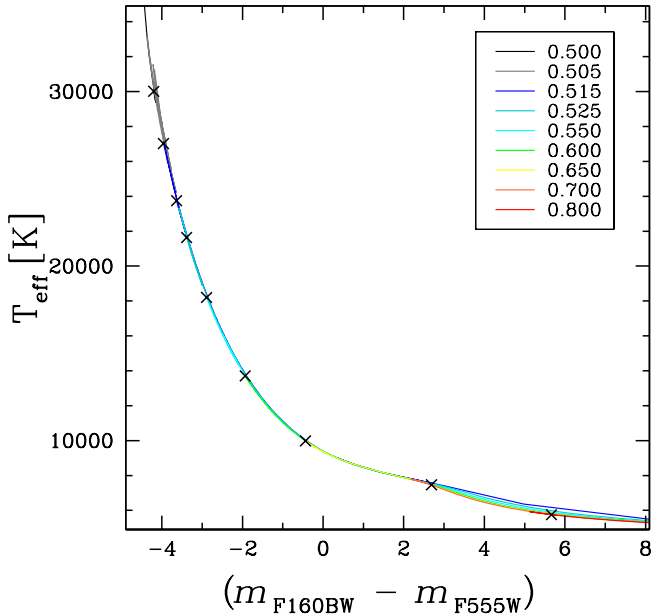} 
\caption{As Fig.~\ref{fig:col-teffZ} but for the ($m_{\rm F160W} - m_{\rm
F555W}$) - \teff\ relations obtained from HB tracks of different masses, with
$\rm {Z} = 0.0003$ and $\rm{Y} = 0.245$. Crosses mark the ZAHB location of each
displayed track (see text for details).} 
\label{fig:col-teffg1}
\end{figure}

\begin{figure} 
\centering
\includegraphics[width=\columnwidth]{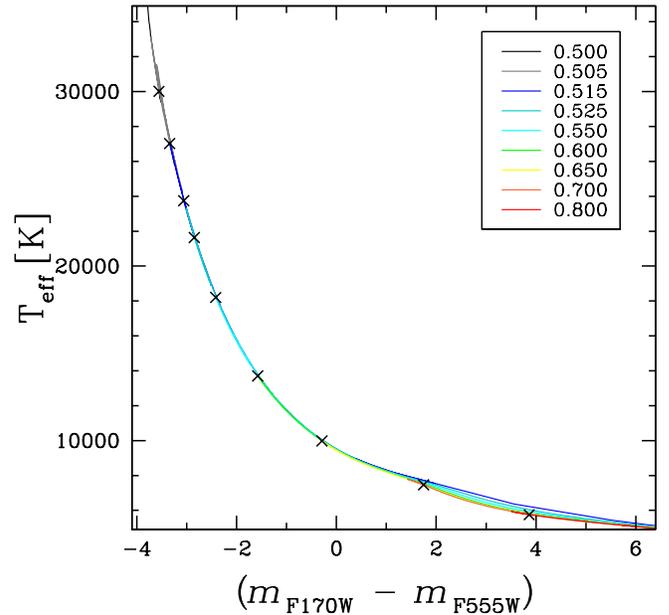} 
\caption{As Fig.~\ref{fig:col-teffg1} but for the ($m_{\rm F170W} - m_{\rm
F555W}$) - \teff\ relations.} 
\label{fig:col-teffg2}
\end{figure}

\begin{figure} 
\centering
\includegraphics[width=\columnwidth]{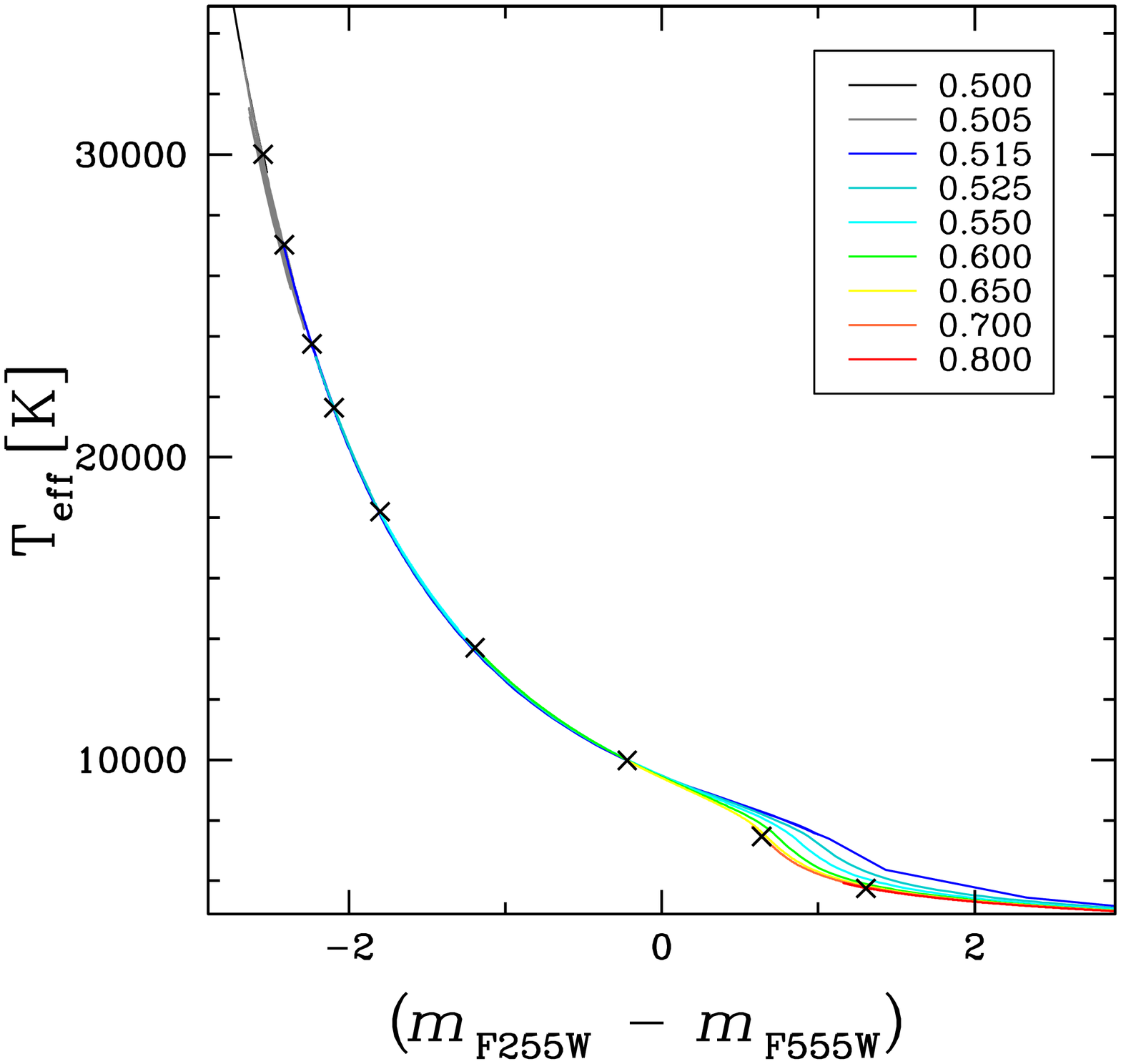} 
\caption{As Fig.~\ref{fig:col-teffg1} but for the $m_{\rm F255W} - m_{\rm
F555W}$ - \teff\ relations.} 
\label{fig:col-teffg3}
\end{figure}

It becomes also clear from Fig. \ref{fig:col-teffZ} that the ($m_{\rm F170W} -
m_{\rm F555W}$) and ($m_{\rm F160W} - m_{\rm F555W}$) colors are the most
sensitive to \teff\ variations (they span a larger range for the same \teff\
interval) at any regime. However, the individual $m_{\rm F160BW}$ and $m_{\rm
F170W}$ magnitudes become very faint for $\teff < 10,000$\,K , and our
photometry in these filters can be severely incomplete.  The second best choice
for temperature determination below 10,000\,K is ($m_{\rm F255W} - m_{\rm
F555W}$); the color range is almost halved, but the completeness of our data is
much more reliable.  Therefore this is our selected color for this temperature
range.  Redder filter combinations are less suited for \teff\ determinations,
and the ($m_{\rm F336W} - m_{\rm F555W}$) color is almost insensitive to \teff\
below $\sim 8,000$\,K.

Given that it is now well known that stars within individual clusters can cover
a non negligible  range of He abundances \citep[see e.g.][\citealt{Mil14} and
references therein]{Dan05,Dal11} we have analyzed the impact of Y variations on
the ZAHB color - \teff\ relations. In particular, Figure \ref{fig:col-teffY}
displays the effect of two different initial He mass fractions ($\rm{Y} = 0.245$
and $\rm{Y} = 0.300$) on the predicted ZAHB color - \teff\ relations for the
filter combinations we are going to use in our investigations.  It emerges
clearly that the effect of the in principle unknown Y content of the observed
stars is completely negligible at any \teff.  Another factor that may affect the
relation is the fact that stars along the observed HB are born with different
patterns of CNONa abundances. In \citet[see appendix A]{Dal13-1} we have already
shown that also the effect of the CNONa anticorrelations is negligible.

After having established the filter combinations to use for our \teff\
estimates, and assessed the effect of the chosen initial chemical composition of
the ZAHB, we address here the issue of the effect of surface gravity. We recall
that we apply a color - \teff\ relation obtained from the ZAHB to the observed
color of HB stars, irrespective of whether they are evolving close to the ZAHB
or are evolved off-ZAHB. To assess the extent of systematic biases introduced by
this procedure, Figures \ref{fig:col-teffg1}, \ref{fig:col-teffg2} and
\ref{fig:col-teffg3} display the relevant color - \teff\ relations for HB
tracks of different masses and the same initial chemical composition ($\rm{Z} =
0.0003, \rm{Y} = 0.245$). For the ($m_{\rm F160W} - m_{\rm F555W}$) and ($m_{\rm
F170W} - m_{\rm F555W}$) colors the effect of off-ZAHB evolution is negligible
at all \teff\ covered by the models. In fact, in the color - \teff\ diagram the
off-ZAHB evolution of the models nicely overlaps with the ZAHB sequence. The
effect is small also for $\teff < 8,000$\,K, but, as anticipated, we are not
going to use these colors combinations for red HB stars.  Also for the ($m_{\rm
F255W} - m_{\rm F555W}$) color diagram evolutionary effects are in general
negligible.  However they start to be visible below $\sim 8,000$\,K. The maximum
effect is at \teff\ in the range between $\sim 6,000$ and 7,500\,K and amounts
to $\sim 1,000$\,K.  This represents the maximum systematic error due to
evolutionary effects for our individual temperature estimates below 10,000\,K.

As discussed in detail by \cite{Gir08} the extinction coefficients for
broadband filters may depend on the effective temperature of the stars. If this
is the case, the theoretical color - \teff\ relations should be modified to take
this into account  when applied to the cluster HB stars. Following the methods
outlined in \cite{Gir08} we have verified that for the F170W filter and $\teff >
10,000$\,K, the extinction displays a negligible dependence on the temperature,
and the same is true for the F255W filter when $\teff < 10,000$\, K.  This
guarantees that the slopes of the color - \teff\ relations that we use do not
depend on the individual values of $\rm{E\,(B-V)}$.

Finally, for the very rare cases (in our sample) of clusters hosting blue-hook
stars, which are hotter than the bluest end of any ZAHB, we will use
hot-flashers models as already done for \ngc2808 \citep{Dal11}.

\section{Observations and data reduction}
\begin{figure} 
\centering
\includegraphics[width=\columnwidth]{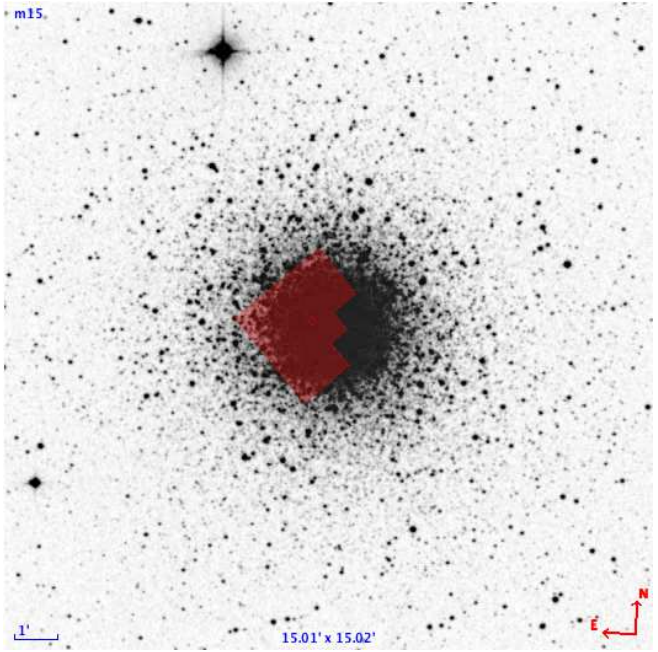} 
\caption{Footprint of the WFPC2 images used in this work, superimposed on a
$15\arcmin \times\,15\arcmin$ DSS POSS-I plate. North is up and East is left.} 
\label{fig:fov}
\end{figure}

The dataset used in this work consists of 20 WFPC2 images collected on Apr 16th
2009 and centered on the core of M15, as shown in Figure~\ref{fig:fov}.
Table~\ref{tab:log} summarizes filters and exposure times of the observations.

\begin{deluxetable}{lc}
\tablewidth{0pt}
\tablecaption{WFPC2 images of M15 used in this work.\label{tab:log}}
\tablehead{\colhead{Filter} & \colhead{No. of images $\times$ Exp. time (sec)}}
\startdata
F160BW & $1 \times 1500$ \\
F170W & $1 \times 1000 + 3 \times 700$ \\
F255W & $3 \times 1200$ \\
F300W & $1 \times 80 + 3 \times 600$ \\ 
F336W & $1 \times 40 + 1 \times 700 + 2 \times 800$ \\ 
F555W & $1 \times 1 + 3 \times 40$
\enddata
\end{deluxetable}

All the images have been pre-processed through the standard \hst/WFPC2 pipeline
for bias - dark subtraction and flat fielding. They have been corrected also for
the pixel area and `34th row' effect, which are specific instrument-induced
signal variations \citep{Bag02}. Single frames were extracted from the WFPC2
mosaic image and, for each given chip and filter, we combined them with the
\textsc{iraf}\,\footnote{\textsc{iraf} (Image Reduction and Analysis Facility)
is distributed by the National Optical Astronomy Observatory, which is operated
by the Association of Universities for Research in Astronomy, Inc., under
cooperative agreement with the National Science Foundation.} task
\texttt{imcombine}, applying a cosmic-ray rejection algorithm.

The photometric analysis was performed both on raw and median images, by means
of the \textsc{daophot\,iv/allstar} suite \citep{Ste87}. We first obtained
preliminary catalogs independently for each frame by using the \textsc{daophot}
star-finding algorithm. Then, we built up an initial common list of
identifications by geometrically matching, with \textsc{daomatch/daomaster}
\citep{Ste93}, these preliminary catalogs.

Cross-identifications are assumed to be real detections if they occur in at
least a minimum number of frames which varies from filter to filter. This step
is aimed at discarding possible spurious cosmic-ray detections survived on the
median image as well as fake detections from the preliminary catalogs of the raw
images. Since we have only one exposure in the F160BW filter, a reliable star
list in this band has been obtained after cross-matching the preliminary F160BW
catalog with those derived in F170W.

Once obtained a reliable common list of objects, we estimated a suitable point
spread function (PSF) on the median image through the selection of bright and
isolated stars uniformly distributed over the entire chip area. Given the
stability of the \hst\ PSF, and since the S/N of the median is higher than that
of the single image, we applied the PSF model obtained for the former to the
latter. We assumed a quadratic variation of the PSF with the position on the
frame for all the images but the F160BW and F170W ones, where the limited number
of stars restricted our choice to a linearly variable PSF model.

For each chip we created two master lists: the first includes stars detected in
at least two median images among those obtained in the F255W, F300W, F336W and
F555W filters, after they have been transformed to a common reference system;
the second includes stars detected in at least two F170W exposures. Stars in the
master lists were then forced to fit in single images with \textsc{allframe}
\citep{Ste94}, separately for each filter. 

Afterwards, the following quantities were computed and applied in sequence to
the new instrumental catalogs: 

\begin{itemize}
\item aperture correction within a nominal infinite aperture. It was obtained by
subtracting 0.1\,mag to the correction for an aperture within a radius of
$0.5\arcsec$ calculated with \textsc{daogrow} \citep{Ste90}. 
\item correction for Charge Transfer Efficiency loss by means of the equations
provided in \cite{Dol09};
\item corrections for UV Contamination and Long-term Quantum Efficiency
Change by reading the value of the \texttt{ZP\_CORR} keyword in the file header.  
\end{itemize}

\begin{figure} 
\centering 
\includegraphics[width=\columnwidth]{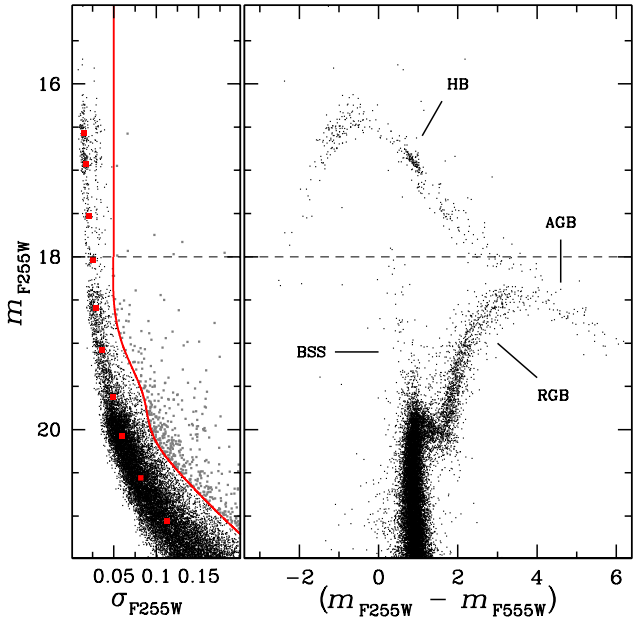} 
\caption{{\it Left}: Plot of magnitude {\it vs.} standard error of magnitude in
the F255W band. The red squares mark the median point in each 0.5\,mag bin. The
red solid line divides `well-measured' (black dots) and rejected (gray dots)
stars. {\it Right}: ($m_{\rm F255W}, m_{\rm F255W} - m_{\rm F555W}$) CMD of the
well-measured stars. The different evolutionary branches have been labeled.}
\label{fig:err} 
\end{figure}

\subsection{Photometric and astrometric calibration}
For each filter we matched the catalog of the single images, obtaining six
different master lists, where the instrumental magnitude and the error of the
magnitude are, respectively, the weighted mean of the single image measurement
reported to the system of the reference frame of the transformation and the
standard error of the mean, based on individual frame standard errors
\citep{Fer91,Fer92}. Instrumental magnitudes were then transformed into the
VEGAMAG system adding the zero points listed in table 5.1 of the WFPC2 Data
Handbook\,\footnote{\url{http://www.stsci.edu/hst/wfpc2/documents/handbook/WFPC2_DHB.html}}
\citep[ver.5.0, July 2010,][]{Bag97}. Since there is no overlap among the
different WFPC2 chips, we adopted an empirical procedure to quantify the zero
point variation among them (see appendix \ref{sec:corr}). The final total
catalog counts $\sim 44,000$ stars.

\begin{figure}
\centering
\includegraphics[width=\columnwidth]{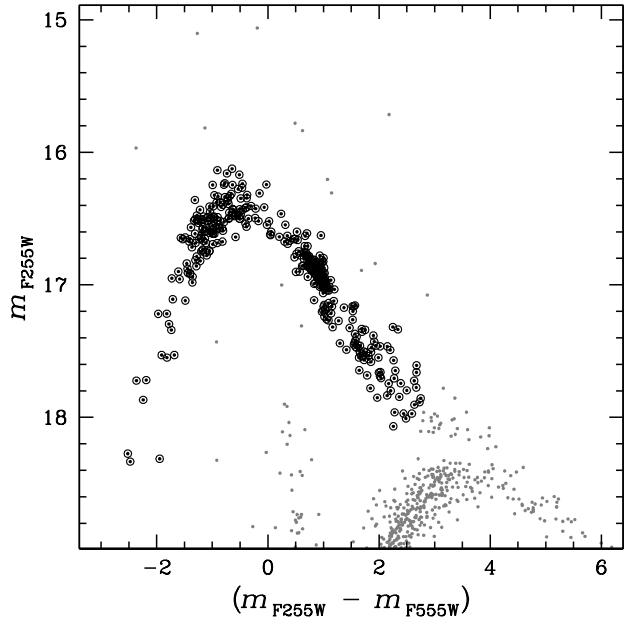}
\caption{Zoom of the HB region in the ($m_{\rm F255W}, m_{\rm F255W} - m_{\rm
F555W}$) CMD of M15. Black circles represent the final selection of HB members.}
\label{fig:var}
\end{figure}

\begin{figure*}
\centering
\includegraphics[angle=270,width=\textwidth]{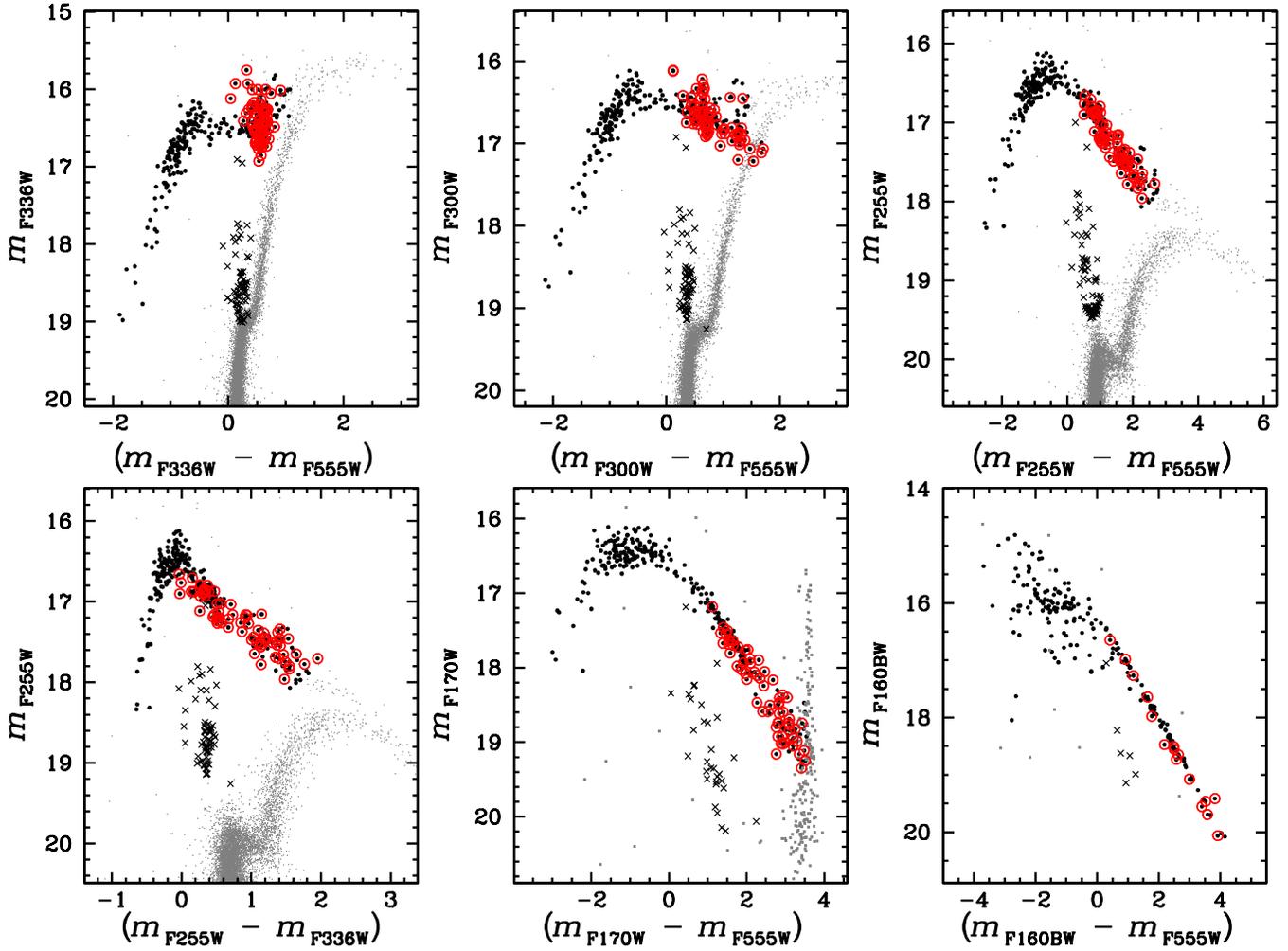}
\caption{CMD of M15 in six different combination of filters. The various stellar
components of the clusters are coded as following: gray dots = MS + SGB +
RGB/AGB; black cross = BSS; black black point = HB. The RR\,Lyr\ae\ in
common between our catalog and the Clement's {\it Catalogue of Variable Stars in
Galactic Globular Clusters} are marked by red circles.}.
\label{fig:6cmd}
\end{figure*}

In order to transform the star instrumental pixel-based position into the
absolute equatorial coordinate system we performed a two-step procedure
\citep[as already done in previous works; see for example][]{Dal09} Firstly, we
reported the instrumental WFPC2 coordinates to the reference system of the M15
ACS catalog, available in the ACS Survey of Galactic Globular Cluster
Database\,\footnote{\url{http://www.astro.ufl.edu/~ata/public_hstgc/databases.html}}
\citep{Sar07}. Secondly, we put these new coordinates onto the \textsc{2mass}
Point Source Catalogue system \citep{Cut03}. Both steps were achieved by means
of the \textsc{cataXcorr/catacomb} package, developed by P.~Montegriffo at the
Bologna Astronomical Observatory (private communication). The program provided
us with a very accurate astrometric solution for the transformation into the
absolute system of ACS, with typical error of $\sim 0.3\arcsec$ in right
ascension and declination, for stars measured in the Planetary Camera (PC) chip,
(whose pointing is located at the very center of the cluster) and of $\sim
0.02\arcsec$ for the stars in the three Wide Field (WF2, WF3, WF4) cameras. The
corrections to the absolute system of 2MASS were determined with an error of
$\sim 0.16\arcsec$ in both the coordinates.

\subsection{Sample selection and RR\,Lyr\ae\ identification}
The definition of a sample of HB stars measured with high photometric accuracy
is fundamental in our analysis. To this purpose, we adopted the following
procedure. We first identified the RR Lyra\ae\ in our catalog and then used the
magnitude errors of the non variable stars to select the best-measured objects.
Finally, we precisely identified the HB members by looking at their position in
several CMDs.

For the sake of convenience, we started our selection procedure in the F255W
band. Indeed, taking a look at the ($m_{\rm F255W}, m_{\rm F255W} - m_{\rm
F555W}$) CMD in the right panel of Figure~\ref{fig:err}, we see that HB stars
are among the brightest objects in this diagram \citep{Fer03}, most of them
attaining F255W magnitudes brighter than $\sim 18$. As expected, when plotting
the magnitude error as a function of the magnitude we see a typical trend (left
panel of Fig.~\ref{fig:err}): the fainter the star, the larger its error.  We
divided the full magnitude range into small intervals of 0.5\,mag and for each
one we computed the median value of the magnitude and of the error (red
squares). Then, in each magnitude bin at $m_{\rm F255W} > 18$, we selected all
stars with an error within $3\sigma$ from the median value and we fit the upper
boundary of this error distribution with a cubic spline. An upper limit of
0.05\,mag was instead adopted for stars brighter than $m_{\rm F255W} = 18$.
Stars with a magnitude error smaller than the fit value were flagged as
`well-measured' (black dots). Since a similar trend for the magnitude error is
found in the other bands, the same selection method was applied to the F170W,
F300W, F336W and F555W magnitudes\,\footnote{A bin of 0.4\,mag was adopted for
the analysis in the F300W, F336W and F555W filters.}. In principle only those
stars satisfying the selection criterion in all the filters should be flagged as
`well-measured' objects. However, since the selection criterion in the F255W
filter excludes most of the stars with uncertain measures also in other filters,
in the following we consider as `well measured' all the objects satisfying only
the selection in the F255W filter.  Figure~\ref{fig:var} presents the final
selection of the well measured HB stars (marked with large empty circles). As
can be seen the considered sample includes the vast majority of the observed HB
stars. Also visible in the figure is the brightest portion of the RGB and AGB
at ($m_{\rm F255W} - m_{\rm F555W} \gtrsim 2$ and the BSS sequence at ($m_{\rm
F255W} - m_{\rm F555W}$) $\sim 0.5$ extending up to $m_{\rm F255W} \sim 18$.
With the adopted selection we may either lose a few genuine HB stars or include
some post-HB objects, but they constitute only a negligible fraction of the
selected sample. In fact, based on the evolutionary time-scales, we expect just
one post-HB star every one hundred genuine HB objects \citep[see,
e.g.,][]{Sch12}.  Figure~\ref{fig:6cmd}, which displays the final CMDs in
different band combinations, shows that the selected stars look like genuine HB
objects also in the other CMDs.  Moreover, the ($m_{\rm F336W}, m_{\rm F336W} -
m_{\rm F555W}$) CMD demonstrates that our choice of the red boundary of the HB
selection is reliable, because in that CMD, AGB stars are better separated from
the HB.

In addition to the HB stars, we highlighted the position of the BSSs in all CMDs
of Fig.~\ref{fig:6cmd}. It is worth to notice that moving from pure optical
($m_{\rm F336W}, m_{\rm F555W}$) to mid UV - optical ($m_{\rm F255W}, m_{\rm
F555W}$) and far UV
- optical ($m_{\rm F160BW}, m_{\rm F555W}$) CMDs, the brightest and reddest
  stars, namely the RGBs/AGBs, become progressively fainter with respect to the
BHBs and EHBs which at the contrary are the brightest objects in the ($m_{\rm
F160BW}, m_{\rm F160BW} - m_{\rm F555W}$) and ($m_{\rm F170W}, m_{\rm F170W} -
m_{\rm F555W}$) CMDs. 

In the latter case (see bottom-middle panel of Fig.~\ref{fig:6cmd}), we
observe two specific features:

\begin{enumerate}
\item the RGB/AGB sequence is vertical and the reddest RGB/AGB stars are as
bright as the BHBs/EHBs; 
\item the reddest cluster HBs merge with the vertical RGB/AGB sequence.  
\end{enumerate}

Both features are caused by the particular behavior of the F170W filter.  This
filter is characterized by a significant {\it red leak} due to a tail of
sensibility of the spectral response curve at long wavelengths. According to
that, cool stars appear much brighter in F170W than in other UV filters
\citep{Gir08}.

Given that RR\,Lyr\ae\ are observed at random phase, they have been excluded
from our HB sample. M15 is known to host about 200 RR\,Lyr\ae\ \citep{Cor08}.
Since our dataset is not suited for variability analysis, RR\,Lyr\ae\ have been
observed at random phase. A list of variables is however available in the {\it
Catalogue of Variable Stars in Galactic Globular Clusters} \citep{Cle01}, from
which we retrieved the position and classification of 168 RR\,Lyr\ae. Among
them, only 82 fall in the surveyed FoV. By using \textsc{cataXcorr} we
geometrically matched our catalog and the Clement's one, finding 72
cross-identifications. Among the stars not in common, four lie in the gaps of
the WFPC2 mosaic and for the remaining six we cannot establish an unambiguous
association with stars in our list. Their position in the various CMDs is
shown in Figure~\ref{fig:6cmd}.

\begin{figure*}
\includegraphics[width=\textwidth]{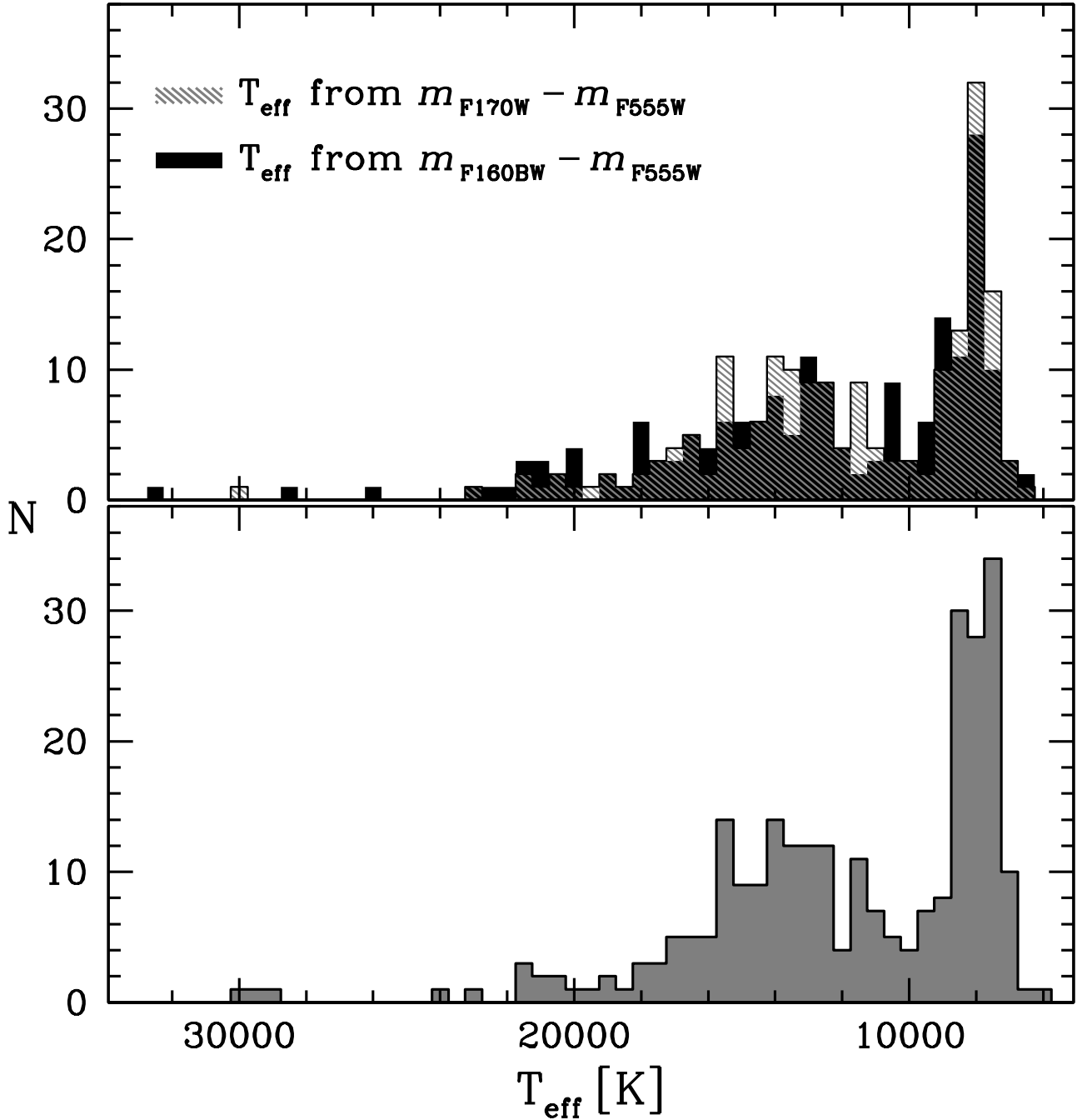}
\caption{Top: Histograms of the temperature distribution of the HB stars in
M15 obtained from the ($m_{\rm F170W} - m_{\rm F555W}$) - \teff\ (solid black) and
($m_{\rm F170W} - m_{\rm F555W}$) - \teff\ (gray shaded) relation for all the HB stars with a
color estimate in the ($m_{\rm F160BW}, m_{\rm F160BW} - m_{\rm F555W}$) CMD.
Bottom: Histogram of the temperature distribution of the HB stars in M15
obtained from the ($m_{\rm F170W}
- m_{\rm F555W}$) - \teff\ relation for stars with $\teff\ > 10,000\,K$, and from
  the ($m_{\rm F255W} - m_{\rm F555W}$) - \teff\ relation for stars with $\teff\ <
10,000$\,K (see text for details). All the distributions have been obtained
adopting a BaSTI ZAHB model with $\feh = -2.14$ and ${\rm Y} = 0.245$.}
\label{fig:hist}
\end{figure*}

\section{Temperature distribution along the HB} \label{sec:distrib}
The typical photometric error of the selected HB stars in the colors based on
F170W and F255W is $\sigma \sim 0.03$, corresponding approximately to a small
random error $\Delta \teff \sim 100$\,K on the individual estimates. A realistic
uncertainty on the cluster reddening, $\Delta\rm{E\,(B-V)}$ by 0.01-0.02\,mag,
would introduce a systematic error of the order of 250-500\,K, that dominates
compared to the random error arising from photometry.  On the other hand the
photometric errors in F160BW are much larger, of the order of $\sigma \sim
0.15$, corresponding to $\Delta \teff \sim 1,000$\,K. As a consequence, even if
in principle the ($m_{\rm F160BW} - m_{\rm F555W}$) color is well suited for our
analysis, in this case it is strongly affected by the uncertainties caused by
photometric errors. However, given that M15 is the only cluster for which a
direct comparison between F160BW and F170W is possible, we decided to use also
the F160BW band with this caveat in mind.

All the color - \teff\ relations used in the following discussion have been
obtained through ZAHB models with metallicity $\feh = -2.14$, which is the
closest value to the metallicity of M15 \citep[$\feh =-2.39 \pm 0.14$;][]{Roe14}
available in the BaSTI database, and primordial helium abundance, namely
${\rm Y} = 0.245$.  We adopted a reddening $\rm{E\,(B-V)} = 0.10$ \citep[][2010
update]{Har96} to correct the theoretical colors.

The top panel in Figure~\ref{fig:hist} displays the derived temperature
distributions obtained from the ($m_{\rm F160BW} - m_{\rm F555W}$) - \teff\
relation (black histogram) and ($m_{\rm F170W} - m_{\rm F555W}$) - \teff\
relation (shaded histogram) for all the HB stars having a color estimate in the
($m_{\rm F160BW}, m_{\rm F160BW} - m_{\rm F555W}$) CMD.  Both show a main peak
located at $\teff \sim 8,000$\,K and a broad distribution extending from $\sim
10,000$\,K to $\sim 22,000$\,K. The three stars hotter than $\sim 26,000$\,K in
the ($m_{\rm F160BW}
- m_{\rm F555W}$) distribution attain lower \teff\ in the ($m_{\rm F170W} -
  m_{\rm F555W}$) temperature distribution. This offset is likely due to the
large photometric errors in ($m_{\rm F160BW} - m_{\rm F555W}$), at these colors.
According to the Kolmogorov-Smirnov test, the distributions of effective
temperatures obtained with the two colors for $\teff < 22,000$\,K are
statistically indistinguishable.

The complete temperature distribution of the HB stars of M15 is shown in
the bottom panel of the Fig~\ref{fig:hist}. It has been obtained through the
combination of F555W with F170W magnitudes, for stars having $\teff >
10,000$\,K according to the ($m_{\rm F170W} - m_{\rm F555W}$) - \teff\ relation,
and through the combination of F555W with F255W magnitudes for stars having
$\teff < 10,000$\,K, according to the ($m_{\rm F255W} - m_{\rm F555W}$) - \teff\
relation (see Sect. \ref{sec:efftemp} for details). This artificial
splitting of the HB sample may generate duplication or loss of a few objects. We
carefully checked the final sample and we found that only three objects would
have been not considered. We therefore assigned them the average value of the
temperature obtained from the two color indices.

\begin{figure}
\includegraphics[width=\columnwidth]{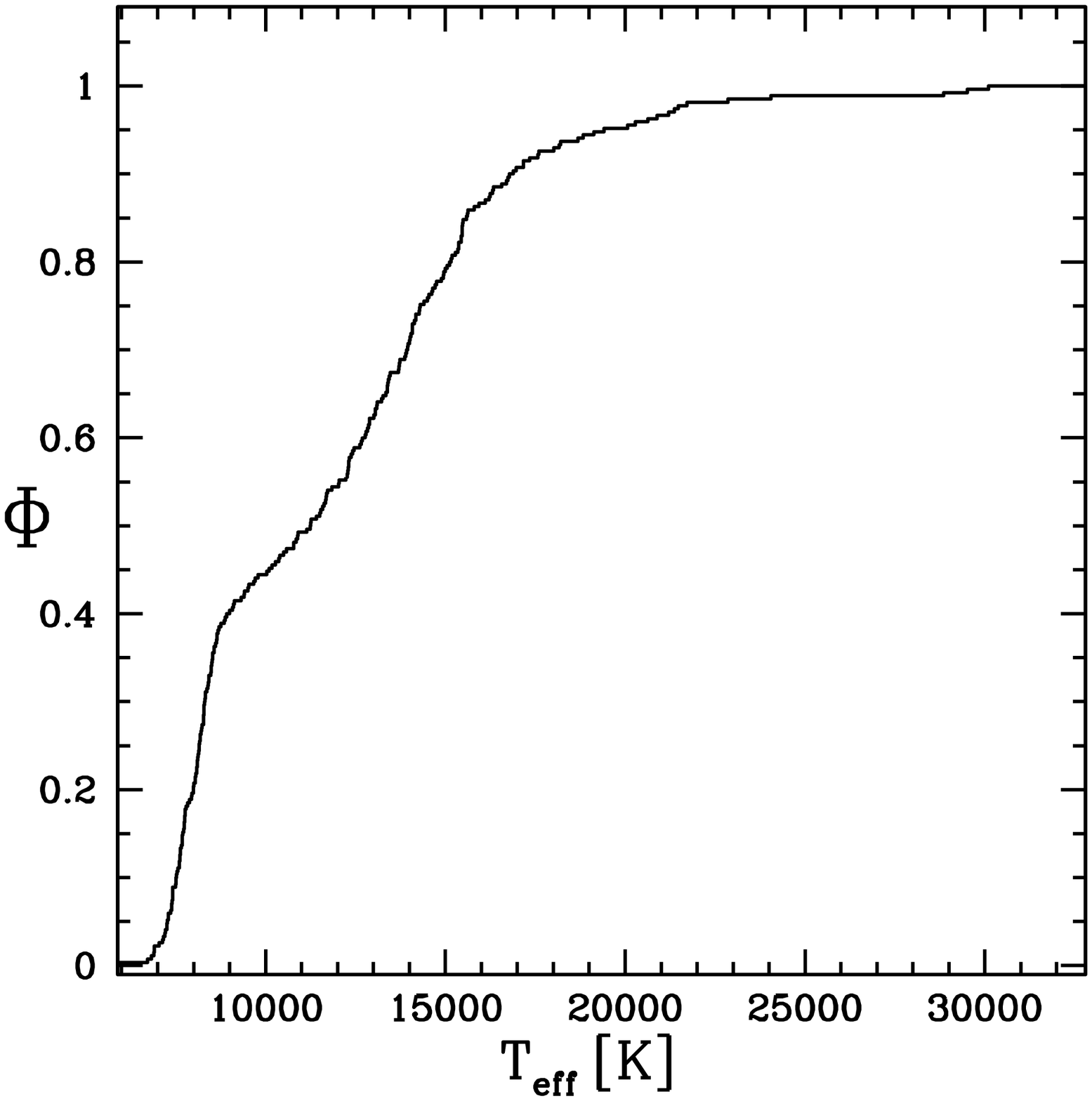}
\caption{Cumulative temperature distribution for the HB stars of M15 obtained
from the ($m_{\rm F170W} - m_{\rm F555W}$) - \teff\ relation and adopting an
$\alpha$-enhanced ZAHB model with $\feh = -2.14$ and ${\rm Y} = 0.245$.} 
\label{fig:ks}
\end{figure}

On the whole we observe a main component peaked at $\sim 8,000$\,K and confined
below $\sim 10,000$\,K.  The second component, vaguely Gaussian is peaked at
$\teff \sim 14,000$\,K. Although the hottest HB star is located at 30,000\,K in
qualitative agreement with what found by \cite{Moh97} and
\cite{Rec06}, the derived distribution shows that only a few stars can be found
at $\teff > 20,000$\,K \citep{Moh95}. This can be more easily appreciated from
Figure~\ref{fig:ks}, which displays the cumulative temperature distribution of
the HB stars in M15 obtained before. As can be seen, essentially the 95\% of the
stars in the M15 HB is below $\teff = 20,000$\,K.

\subsection{Comparison with spectroscopic data} 
The temperature of HB stars obtained through the color - \teff\ relation might
be prone to systematic errors introduced by the theoretical model assumptions
and the parameters (e.g. reddening) used to fit the observations . In order to
validate the temperature estimates, spectroscopic temperature determinations of
the same HB stars can be used. 

In the case of M15, spectroscopic observations of the cluster's HB stars were
performed by \cite{Moh95,Moh97} and by \cite{Beh03}. The three spectroscopic
samples were selected from the ground-based U, B, V catalog of \cite{Buo83},
which unfortunately does not overlap our FoV. Therefore, no direct comparison
has been possible.  However, we devised the following procedure to test the
reliability of our temperature scale. 

\begin{figure*}
\includegraphics[width=\textwidth]{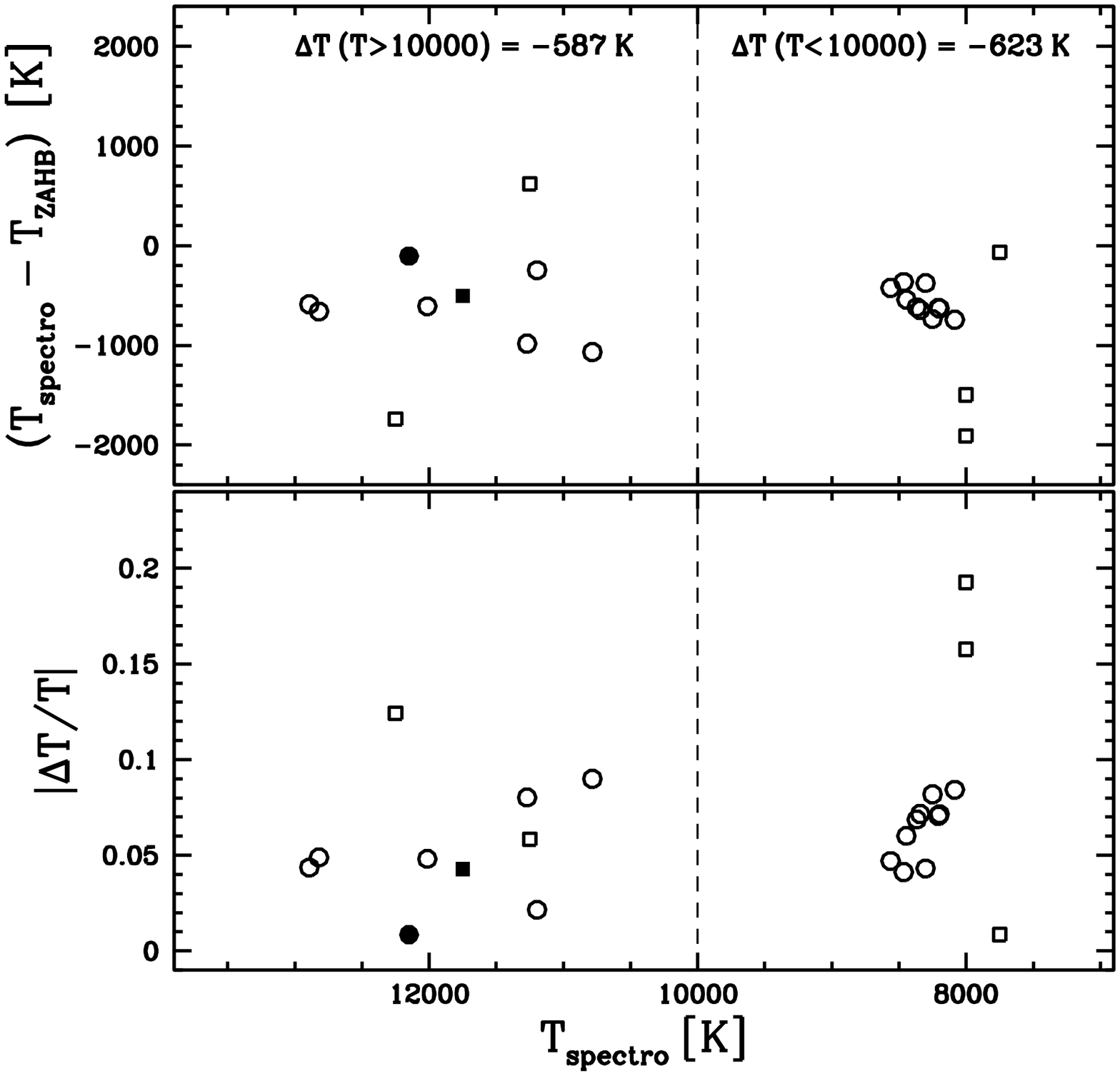}
\caption{{\it Top}: Difference between spectroscopic ($\rm T_{spectro}$) and
photometric ($\rm T_{ZAHB}$) temperature estimates for the HB stars in M15.
Squares correspond to spectroscopic measurements of \cite{Moh95}, circles to
those of \citep{Beh03}. The filled symbols marks the only star \citep[B348 in
][]{Buo83} in common between the two samples. The median difference value for
the stars with ${\rm T_{spectro}} < 10,000$\,K and ${\rm T_{spectro}} >
10,000$\,K are also labelled. {\it Bottom}: Absolute value of the relative
temperature difference for the same stars.} 
\label{fig:deltaT}
\end{figure*}

Since the photometry of \cite{Buo83} is fully compatible with that of
\cite{Ste94}~\footnote{Available on the Peter Stetson's Photometric Standard
Fields website:
\url{http://www3.cadc-ccda.hia-iha.nrc-cnrc.gc.ca/community/STETSON/standards/}},
we cross-matched the latter dataset with our catalog, finding 136 stars in
common, 26 being HB stars, mainly populating the red portion of the branch. We
used this sub-sample of stars in order to find a relation between the $\rm (B
-V)$  and the ($m_{\rm F170W} - m_{\rm F555W}$) colors. This relation has been
used to transform the $\rm (B -V)$ color of 17 stars measured by \cite{Beh03}
and 6 stars observed by \cite{Moh95}, in the UV color and then in temperature.
This finally allowed us to compare the temperatures obtained from the ZAHB
colors with those obtained from the spectroscopy. The results are shown in
Figure~\ref{fig:deltaT}. The top panel shows the difference between the
spectroscopic temperature estimates and the temperatures extracted from the
color - \teff\ relation of the adopted ZAHB model (see Sect.~\ref{sec:distrib}).
We found a median difference of $\approx -600 \pm 20$\,K, both below and above
$10,000$\,K.  Such a difference is comparable to the average precision of our HB
temperature estimates. Although the difference is systematic, it can be
easily explained by just a 0.02 mag overstimate of the cluster reddening (see
Sect.~\ref{sec:distrib}). Note that one star \citep[B348 in ][]{Buo83} is in
common between the two spectroscopic catalogs. Interestingly the difference
between the two spectroscopic temperatures ($\sim 400$\,K) is similar to the
difference with respect to our photometric measures (see the two filled symbols
in Fig.~\ref{fig:deltaT}). In the lower panel of Fig.~\ref{fig:deltaT} the
absolute value of the relative difference is plotted: all the stars scatter
around $|\Delta\,{\rm T}/{\rm T}| = 0.05$ with only three objects above 0.10.
The results of this comparison clearly confirms the reliability of our approach.

\section{Summary and conclusions}
This work is part of a series of papers that aims at studying the temperature
distribution of a large sample of GGCs, for which we obtained homogeneous UV and
optical \hst\ data.

The temperature is derived by using the ZAHB color - \teff\ relations by
\cite{Pie06}. We investigated here the most suitable colors for temperature
estimates, the effect of metallicity variations, the impact of neglecting
off-ZAHB evolution and also the impact of neglecting differences of He
abundances among HB stars on the adopted color - \teff\ relations. 

As already shown by \cite{Roo08} and \cite{Dal11,Dal13-1} we found that the
($m_{\rm F160BW} - m_{\rm F555W}$) and ($m_{\rm F170W} - m_{\rm F555W}$) color
combinations are most sensitive to \teff\ variations. However since the sample
of red HB stars can be partially incomplete in far-UV bands for some clusters,
the use of the ($m_{\rm F255W} - m_{\rm F555W}$) can be considered as a good
alternative choice at $\teff < 10,000$\,K.  We have shown that variations in
metallicity and He have a negligible effect on the ZAHB color - \teff\
relations. Moreover we have found that also the effect of gravity due to the
off-ZAHB evolution of HB stars has an almost negligible effect in any color
combination for $\teff\ > 8,000$\,K, while at lower temperatures in the ($m_{\rm
F255W} - m_{\rm F555W}$) color, it introduces a maximum systematic error of
$\sim 15\%$ on individual temperatures.  These results fully justify the use of
the ZAHB color - \teff\ relations.

As a test case in our analysis, we have applied our approach to the metal poor
GGC M15. We performed a photometric analysis of 20 images centered on the core
of the cluster, collected in the filters F160BW, F170W, F255W, F300W, F336W and
F555W.  Since M15 is the only cluster in our survey for which we have a total
filter coverage, it serves as calibrator for our database. 

We derived the temperature of M15 HB stars by using the theoretical color -
\teff\ relation for $\feh = -2.14$. We compared the temperature
distributions from the ($m_{\rm F160BW} - m_{\rm F555W}$) and ($m_{\rm F170W} -
m_{\rm F555W}$) colors, finding a good agreement between the two. This outcome
enables a direct comparison with results already published by our group for
clusters with only F160BW observations. However, since the other clusters in our
sample have been observed only in the F170W filter, we adopted the temperature
distribution obtained from the ($m_{\rm F170W} - m_{\rm F555W}$) - \teff\
relation. Moreover, we adopted the temperatures obtained by the ($m_{\rm F255W}
- m_{\rm F555W}$) - \teff\ for $\teff < 10,000$\,K because the completeness of
  our data is higher in the F255W band for relatively cold stars. 

The temperature distribution of the HB stars in M15 is clearly multimodal with a
main component peaked at $\sim 8,000$\,K and confined below 10,000\,K. The
second component is peaked at $\teff \sim 14,000$\,K and it extends up to $\teff
\sim 20,000$\,K.  Indeed even if the hottest HB star in M15 has been found at
$\teff \sim 30,000$\,K, in agreement with \cite{Rec06} and \cite{Moh95}, it is
important to emphasize that 95\% of the HB population in M15 lies below $\teff
\sim 20,000$\,K. We have also shown that the temperature estimates obtained from
ZAHB colors are consistent with previous spectroscopic estimates
\citep{Moh95,Beh03}.

\begin{acknowledgements}
This research is part of the project COSMIC-LAB (\url{http://www.cosmic-lab.eu})
funded by the European Research Council (under contract ERC-2010-AdG-267675).
\end{acknowledgements}


\appendix
\section{Correction for zero point variation of the WFPC2 mosaic}\label{sec:corr}
As specified in the WFPC2 Data Handbook, the VEGAMAG zero points are referred to
the last instrument calibration, which was performed as early as in 2002.  Since
the observations presented here were acquired in 2009 (7 years after) the
calibrations could be out of date and could not provide anymore a precise
calibration of each WFPC2 chip. In order to verify the validity of the adopted
calibration we decided to plot different color-magnitude diagrams, using a
different color code for stars belonging to the four different chips of the
WFPC2 mosaic. In all of them we observed a clear misalignment of the
evolutionary sequences both in color and magnitude. Therefore, we devised an
empirical procedure for the re-alignment of the magnitudes of the four chips,
using as reference the WF3 photometry as our reference catalog.

\begin{figure} 
\centering
\includegraphics[angle=270,width=\columnwidth]{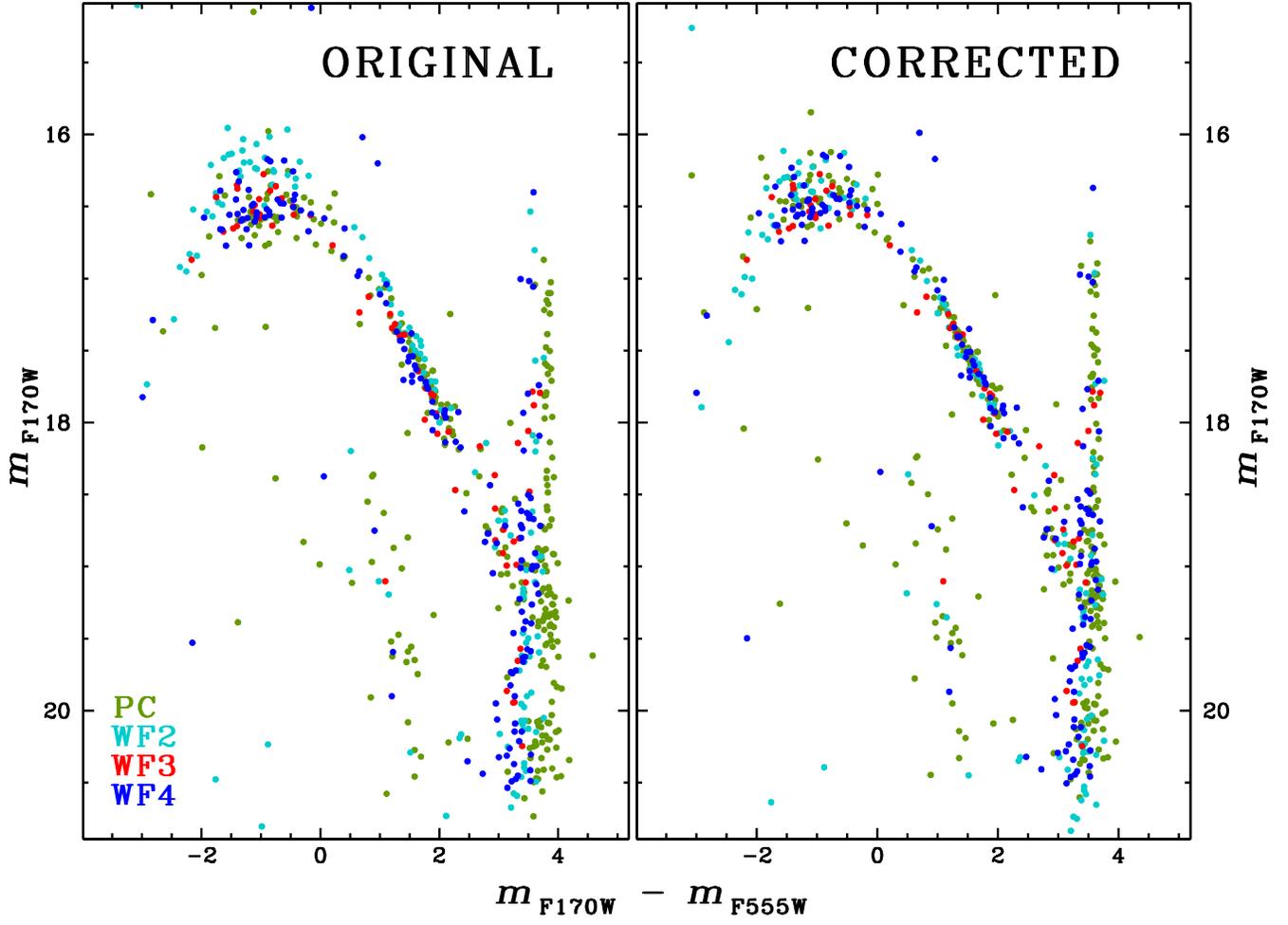} 
\caption{($m_{\rm F170W}, m_{\rm F170W} - m_{\rm F555W}$) CMD of M15, showing
with different colors the stars detected in the PC, WF2, WF3 and WF4 chips of
WFPC2. {\it Left}: CMD obtained by adopting the zero points listed in the WFPC2
Data Handbook fot the calibration of instrumental magnitudes to the VEGAMAG
system. {\it Right}: CMD obtained after the PC, WF2 and WF4 magnitudes were
manually shifted to match the WF3 (red points) sequence.}
\label{fig:170} 
\end{figure}

We started by fixing the F170W and the F555W magnitude displacements by
re-aligning the RGB/AGB and HB sequences observed in the ($m_{\rm F170W}, m_{\rm
F170W} - m_{\rm F555W}$) and ($m_{\rm F555W}, m_{\rm F170W} - m_{\rm F555W}$)
CMDs (see Figure~\ref{fig:170}).  Indeed, in these diagrams, the RGB/AGB
sequence appears as an almost perfect vertical strip of stars, while the red HB
appears as a sharp, diagonal sequence in the ($m_{\rm F170W}, m_{\rm F170W} -
m_{\rm F555W}$) CMD and horizontal in the ($m_{\rm F555W}, m_{\rm F170W} -
m_{\rm F555W}$) one \citep{Lag14}. Hence we were able to determine the
corrections to the adopted zero points in the F170W and F555W filters, by
shifting in color and/or magnitude the RGB/AGB and HB sequences of the PC, WF2
and WF4 in order to match those obtained for the WF3 catalog. 

After that, the displacement in $m_{\rm F255W}, m_{\rm F300W}, m_{\rm F336W}$
were straightforwardly recovered keeping the $m_{\rm F555W}$ shift fixed and
assuming that the displacement observed in the ($m_i, m_i - m_{\rm F555W}$) CMD,
with i = F255W, F300W, F336W, was entirely due to the $i$-th filter.  

In the case of the F160BW magnitudes (because of the large calibration
uncertainty) we adopted the best fit ZAHB model in the ($m_{\rm F160BW}, m_{\rm
F160BW} - m_{\rm F555W}$) CMD as reference system and the magnitudes obtained for
all the WFPC2 chips have bee reported to that reference.
\end{document}